\documentstyle[12pt]{article}
\begin{document}

\baselineskip 0.86cm

\title{On parameters of the Levi-Civita solution}
\author{ A.Z. Wang \thanks{e-mail address: wang@symbcomp.uerj.br,
wang@on.br} \ and  M.F.A. da Silva \thanks{e-mail address: 
 mfas@symbcomp.uerj.br, mfas@on.br} \\
\small Departamento de F\' {\i}sica Te\' orica,
Universidade do Estado do Rio de Janeiro, \\
\small Rua S\~ ao Francisco Xavier 524, Maracan\~ a,
20550-013 Rio de Janeiro~--~RJ, Brazil\\
\small and\\
\small Departamento de Astrof\'{\i}sica, Observat\'orio
Nacional~--~CNPq, \\ 
\small Rua General Jos\'e Cristino 77, S\~ao Crist\'ov\~ao,
20921-400 Rio de Janeiro~--~RJ, Brazil\\
   \\
N. O. Santos \thanks{e-mail address: nos@on.br}\\
\small Universit\'e Paris VI, CNRS$/$URA 769, Cosmologie 
et Gravitation Relativites,\\
\small Tour 22-12, 4\'eme \'etage, Bo\^ite 142, 4, 
Place Jussieu, 75005 Paris, France\\
\small and \\
\small Departamento de Astrof\'{\i}sica, Observat\'orio
Nacional~--~CNPq, \\  
\small Rua General Jos\'e Cristino 77, S\~ao Crist\'ov\~ao,
20921-400 Rio de Janeiro~--~RJ, Brazil} 

\maketitle

\newpage

\begin{abstract}

\baselineskip 0.86cm

The Levi-Civita (LC) solution is matched to a cylindrical shell of an
anisotropic fluid. The fluid satisfies the energy conditions when the
mass parameter $\sigma$ is in the range $0 \le \sigma \le 1$. The mass
per unit length of the shell is given explicitly in terms of $\sigma$,
which has a finite maximum. The relevance of the results to the
non-existence of horizons in the LC solution and to gauge cosmic
strings is pointed out.

\end{abstract}

\vspace{.3cm}

\noindent{PACS numbers: 04.20.Jb, 04.20.Cv, 98.80.Cq, 97.60.Lf }

\newpage

\baselineskip 0.96cm

\section{Introduction}

The number of exact solutions of the Einstein field equations has been
increasing dramatically, especially after the discovery of the so-called
soliton techniques in the late seventies \cite{MC1980}. However, there
are very few of them with physical interpretations \cite{Bonnor1992}.
The understanding of their physics is closely related to the
understanding of their sources.  In fact, it is exactly in this vein
that the work of finding a source for the Kerr vacuum solution still
continues, and so far no success is claimed, although the Kerr
solution was found more than thirty years ago. 

Another rather embarrassing case is the solution found by Levi-Civita
in 1919 \cite{LC1919}. The Levi-Civita (LC) solution is usually
written in the form
\begin{equation}
ds^2 = R^{4\sigma} dT^{2} - R^{4\sigma(2\sigma - 1)}(dR^{2} + dZ^{2}) 
- C^{-2}R^{2 - 4\sigma} d\varphi^2,
\end{equation}
where $\sigma$ and $C$ are two arbitrary constants, and $\{x^{\mu}\} =
\{T, R, Z,\varphi\}$ are the usual cylindrical coordinates with $ -
\infty < T, Z < + \infty, \;\;\; R \ge 0, \;\;\; 0 \le \varphi \le
2\pi$, and the hypersurfaces $\varphi = 0, 2\pi$ being identified.
Except for the cases $\sigma = - \frac{1}{2}, 0, \frac{1}{4},
\frac{1}{2}$, the metric is Petrov type I \cite{GH1969,BM1991}, and has
three Killing vectors, $\xi_{(0)}^{\mu} = \delta^{\mu}_{T},
\xi_{(2)}^{\mu} = \delta^{\mu}_{Z}$, and $\xi_{(3)}^{\mu} =
\delta^{\mu}_{\varphi}$. For $\sigma = - \frac{1}{2}, \frac{1}{4}$, the
metric is Petrov type D, and there is a fourth Killing vector,
$\xi_{(-1/2)}^{\mu} = \varphi\delta^{\mu}_{Z} - Z
\delta^{\mu}_{\varphi}$, and $\xi_{(1/4)}^{\mu} =
\varphi\delta^{\mu}_{T} - T \delta^{\mu}_{\varphi}$, respectively.  The
Killing vector $\xi_{(-1/2)}^{\mu}$ corresponds to a rotation in the
$(Z, \varphi)-$plane, and the corresponding solution is locally
isometric to Taub's plane solution \cite{Taub1951}. This property
suggests to some authors \cite{JK1994} that Taub's solution
 has in fact cylindrical symmetry. However, we intend to take the
opposite view, since the intrinsic curvature of the $(Z, \varphi)-$plane
is identically zero, and we consider this as the main criterion for plane
symmetry.  The Killing vector $\xi_{(1/4)}^{\mu}$ corresponds to a
Lorentz boost. For $\sigma = 0, \frac{1}{2}$, the metric is
locally flat, and in the latter case it is written in a uniformly
accelerated system of coordinates. Except for the last two cases, the
spacetime is asymptotically flat in the radial direction and singular
on the axis. This can be seen, for example, from the Kretschmann scalar
$$
{\cal R} \equiv R_{\alpha \beta \gamma \delta} R^{\alpha \beta 
\gamma \delta} = \frac{64 A \sigma^{2}(2\sigma - 1)^{2}}{R^{4A}},
$$
where $A \equiv 4\sigma^{2} - 2\sigma + 1$. The singularity is usually
believed to represent a cylindrical source, and in realistic models it
should be replaced by a regular region filled with matter. However,
this kind of explanation is unsatisfactory, because it leaves open the
question of whether  or not the matter can be realistic. Considering the
Newtonian limit and time-like geodesics, Gautreau and Hoffman
\cite{GH1969} found that the above interpretation holds only for $0 \le
\sigma < \frac{1}{4}$, and the parameter $\sigma$ represents the mass
per unit length. This conclusion was later confirmed by constructing
explicit sources for the LC solution
\cite{Bonnor1979,LO1980,Fatima1995}, and was recently extended to $0
\le \sigma < \frac{1}{2}$ \cite{BD1992,P1995}.  It is believed that
$\sigma = \frac{1}{2}$ is the maximal value allowed to LC solution such
that it can be interpreted as representing the gravitational field
produced by a cylindrical source \footnote{In \cite{P1995}, a particular
source for the LC solution with $\sigma = 1$ was found, and the source
is composed of ``rather bizarre relativistic material." Yet, in
\cite{GT1987b} it was remarked that when imposing the strong energy
condition, the parameter $\sigma$ has to be in the range $0 \le \sigma
\le 1$, but no detail was given.}. In any case, when $\sigma$ takes
the latter range, a fundamental question arises: in what form is the
parameter $\sigma$ related to the mass per unit length of the
source, since $\sigma = 0, \frac{1}{2}$ all correspond to a spacetime
that is locally flat? Bonnor and Martins \cite{BM1991} proposed a particular
relation for the range $0 \le \sigma \le \frac{1}{4}$, which is clearly
not applicable beyond it.

On the other hand, the LC solution does not possess any horizons. If
our present understanding about the formation of black holes is
correct, this seems to indicate that there is an upper limit to the
allowed linear mass densities of these cylinders, and that this limit
is always below the critical linear mass, above which  horizons are
expected to be formed \cite{LO1980}.

In this paper, we shall address the above mentioned problems. In
particular, by considering more general sources to the LC vacuum
solution, we shall show that all the gravitational fields of the LC
solution with $0 \le \sigma \le 1$ are  produced by physically
reasonable cylindrical sources, and that the mass per unit length has
a finite maximum at $\sigma = \frac{1}{2}$. Specifically, the paper is
organized as follows: in section 2, the matching of the LC solutions to
that of Minkowski is given and the energy conditions of the resulting
shell on the matching hypersurface are considered; in section 3 the
various definitions of the mass per unit length of the shell are
considered, while in section 4 our main conclusions are presented.

\section{Matching the LC solutions to that of Minkowski}

In the previous investigations, specific equations of state of matter
fields  were usually assumed. Thus all the results obtained so far seem
model-dependent. In order to avoid this problem, we model the cylinder
source as an infinitely thin cylindrical shell with a finite radius,
and inside the shell the spacetime is assumed to be Minkowski.
Although here we do not have a theorem similar to that of Birkhoff in
the spherically symmetric case \cite{MC1980}, it is quite reasonable to
assume that the source is static,  since the LC solution is static.
Therefore,  our assumption that inside the static shell the spacetime
is Minkowski has no loss of generality, since it is the only static
spacetime that has zero mass density distribution \cite{AT1992}. By
matching Minkowski geometry to that of LC,  in general, we obtain  a
massive shell on the matching hypersurface. To obtain the most general
shell, we require that the metric coefficients be only continuous
across the shell, the minimal requirement in order to have the Einstein
field equations meaningful \cite{GT1987}. Then, using the  formulae
given by Taub \cite{Taub1980}, we calculate the surface energy-momentum
tensor \footnote{ The formulae given by Taub are equivalent to those of
Israel \cite{Israel1966}, when the matching hypersurface is spacelike
or timelike, as shown by Taub himself.}.   Of course, matter shells
such obtained are not always physically realistic, unless  we further
impose the energy conditions \cite{HE1973}.  Thin shells of matter
might be formed in the early stages of the Universe, and may provide
the necessary perturbations to the formations of galaxies and the
large-scale structure of the Universe \cite{Vilenkin1985}.

Before going into the details, we ought to mention the other physical
parameter $C$. Marder \cite{Marder1958} first realized that the LC
solution has two physically independent constants, $\sigma$ and $C$,
while Bonnor \cite{Bonnor1979} was the one who first related the
constant $C$ to an angular defect. Recently, this angular defect was
further interpreted as representing gauge cosmic strings
\cite{Vilenkin1981,Fatima1995}, being objects formed in the
early Universe \cite{Vilenkin1985}.

To get the most general matching, we first make the following
coordinate transformations
\begin{equation}
t = \alpha T, \;\;\;\;\;\; r = A^{-1}R^{A} + a,\;\;\;\;\;\;
z = \beta Z,
\end{equation}
where $\alpha, \beta$ and $a$ are arbitrary constants. Clearly, the
above transformations leave the ranges of the coordinates unchanged.
Then, in terms of the new coordinates $\{x^{\mu}\} \equiv \{t, r, z,
\varphi\}$, the metric (1) takes the Gaussian form
\begin{equation}
ds^2 = \alpha^{-2} B^{4\sigma/A}dt^{2} -  dr^{2} 
- \beta^{-2} B^{4\sigma(2\sigma - 1)/A}dz^{2}   
- C^{-2}B^{2(1 - 2\sigma)/A} d\varphi^2,  
\end{equation}
where $B \equiv A(r - a)$. In this new system of coordinates, the
spacetime singularity at $R = 0$ is mapped to $r = a$. In the
following, we shall assume that the matter shell is located on the
hypersurface $r = r_{0}$, where $r_{0}$ is a constant and that the LC
solution (3) describes the region  $r \ge r_{0}$. To avoid spacetime
singularities outside the shell we require $ r_{0} > a$.  Inside the
shell ($0 \le r \le r_{0}$), the spacetime is Minkowski and the metric
takes the form
\begin{equation}
ds^2 = dt^{2} -  dr^{2}  -  dz^{2}  - r^{2} d\varphi^2.
\end{equation}
To have the Einstein field equations meaningful on $r = r_{0}$, we
impose the first junction conditions
$\left.g_{\mu\nu}\right|_{r=r_{0}^{+}} =
 \left.g_{\mu\nu}\right|_{r=r_{0}^{-}}$. Clearly, for $\mu,\nu = 0, 2$,
 these conditions can be always satisfied by properly choosing the two
arbitrary constants $\alpha$ and $\beta$, while for $\mu,\nu = 3$ they
become
\begin{equation}
r_{0}^{2} = C^{- 2}\left[A(r_{0} - a)\right]^{2(1 - 2\sigma)/A}.
\end{equation}
With the above requirement, we can see that the first derivatives of
the metric coefficients with respect to $r$ are, in general,
discontinuous across the hypersurface $r = r_{0}$, which gives rise to
thin shells \cite{Taub1980}. 

Following Taub (see also \cite{LW1995}), we first introduce a 
tensor $b_{\mu\nu}$ via the relations
\begin{equation}
\left[g_{\mu\nu, \lambda}\right]^{-} = \left. 
g^{+}_{\mu\nu, \lambda}\right|_{r = r_{0}} -
\left. g^{-}_{\mu\nu, \lambda}\right|_{r = r_{0}} = 
n_{\lambda}b_{\mu\nu},
\end{equation}
where $n_{\lambda}$ is the normal vector to the hypersurface $r =
r_{0}$, and is now given by $n_{\lambda} = \delta^{r}_{\lambda}$. Once
$b_{\mu\nu}$ is known, the surface energy-momentum tensor
$\tau_{\mu\nu}$ can be read off from the expression
\begin{equation}
\tau_{\mu\nu} = \frac{1}{2}\left\{ b(n g_{\mu\nu} - n_{\mu}n_{\nu}) + 
n_{\lambda}(n_{\mu}b^{\lambda}_{\nu} + n_{\nu}b^{\lambda}_{\mu}) - 
(n b_{\mu\nu} + n_{\lambda}n_{\delta} b^{\lambda 
\delta}g_{\mu\nu})\right\},  
\end{equation}
where $n \equiv n_{\lambda}n^{\lambda}$ and $b \equiv
b_{\lambda}^{\lambda}$.  In terms of $\tau_{\mu\nu}$ the
energy-momentum tensor $T_{\mu\nu}$ of the four-dimensional spacetime
 takes the form $T_{\mu\nu} = \tau_{\mu\nu}\delta(r - r_{0})$, where
$\delta(r - r_{0})$ denotes the Dirac delta function.

>From Eqs.(3), (4) and (6), it can be shown that the non-vanishing
components of $b_{\mu\nu}$ now are given by
\begin{eqnarray}
b_{0 0} &=& 4\sigma\left[A(r_{0} - a)\right]^{(4\sigma - A)/A},\;\;\;
b_{22} = 4\sigma (2\sigma - 1)\left[A(r_{0} - a)
\right]^{[4\sigma(2\sigma - 1) - A]/A},\nonumber\\ 
b_{33} &=& 2 r_{0} - \frac{2(1 - 2\sigma)}{C}\left[A(r_{0} - a)
\right]^{[2(1 - 2\sigma) - A]/A}.
\end{eqnarray}
Substituting the above expressions into Eq.(7), we find that the 
surface energy-momentum tensor can be written as
\begin{equation}
\tau_{\mu\nu} =  \rho t_{\mu} t_{\nu} + p_{z} z_{\mu} z_{\nu}
+ p_{\varphi} \varphi_{\mu} \varphi_{\nu},
\end{equation}
where  $t_{\mu}= \delta_{\mu}^{0}, z_{\mu} = \delta_{\mu}^{2}$,
$\varphi_{\mu} = r_{0} \delta_{\mu}^{3}$, and 
\begin{equation}
\rho = \frac{ 2 \sigma r_{0} - a A}{A r_{0}(r_{0} - a)}, \;\;
p_{z} = \frac{ 2 \sigma r_{0}(1 - 
2 \sigma r_{0}) - a A}{A r_{0}(r_{0} - a)},\;\;
p_{\varphi} = \frac{4 \sigma^{2}}{A r_{0}(r_{0} - a)}. 
\end{equation}
In writing the above equations we have chosen units such that $G =
1 = c$, where $G$ denotes the gravitational constant and $c$ the speed
of light. 

The above expressions show that the shell consists of an
anisotropic fluid with surface energy density $\rho$ and
principal pressures $p_{z}$ and $p_{\varphi}$ in the $z$- and
$\varphi$-directions, respectively. Clearly this interpretation is valid
only provided that the fluid satisfies certain energy conditions
\cite{HE1973}. Using the above expressions it is easy to show that all
the three energy conditions,   weak, strong and dominant, are satisfied
for $0 \le \sigma \le 1$, by appropriately choosing the constant $a$.  In
particular, when $\sigma = 0$, the solution gives rise to a cosmic string
 with a finite radius  \cite{Vilenkin1981},
\begin{equation}
\rho = - p_{z}  = - \frac{a}{r_{0}(r_{0} - a)} = \frac{1}{r_{0}}\left(1 - 
\frac{1}{C}\right),\;\;\;\; 
p_{\varphi} = 0.
\end{equation}

When $a = 0$, Eq.(10) reduces to
\begin{equation}
\rho = \frac{\sigma}{4\pi A r_{0}}, \;\;\;\;\;\; 
p_{z} = \frac{\sigma(1 - 2\sigma)}{4\pi A r_{0}},  \;\;\;\;\;\;
p_{\varphi} = \frac{\sigma^{2}}{2\pi A r_{0}}.
\end{equation}
It can be shown too that in this case the weak and strong energy
conditions are satisfied for $0 \le \sigma \le 1$, while the dominant
energy condition is satisfied for $0 \le \sigma \le 1/2$. If we further
set $\sigma = 0$, we shall find that no cosmic string now exists.  This
is because in this case the first junction condition (5) requires $C =
1$, which means no angular defect. Thus, to obtain a cosmic string
solution for the case  $\sigma = 0$, it is necessary to have $a \not=
0$.

However, since in this paper we are mainly interested in the physical
meaning of $\sigma$,  in the following section we shall, without loss
of generality, restrict ourselves only to this case. It can be
 shown that the main conclusions obtained there also hold in the
general case.

\section{The mass per unit length of the shell}

There exist various definitions of the mass per unit length of a
cylinder \cite{Marder1958,Israel1977,VW1977}. Here we are mainly
interested in the one given by Vishveshwara and Winicour (VW)
\cite{VW1977}, since it always gives the correct Newtonian limit and
seems more suitable for the present case. The VW definition is based on
the time translation and rotational Killing vectors, which now are
$\xi^{\mu}_{(0)} = \delta^{\mu}_{t}$, and $\xi^{\mu}_{(3)} =
\delta^{\mu}_{\varphi}$, respectively.  In terms of these vectors, the
mass  per unit length $\mu$ of a cylinder is defined by
\begin{equation}
\mu = - \frac{1}{2 \tau}(\lambda_{33} \lambda_{00, \tau} -  
\lambda_{03} \lambda_{03, \tau}),
\end{equation}
where
\begin{eqnarray}
\lambda_{00} &=& \xi^{\nu}_{(0)}\xi_{\nu (0)},\;\;\; 
\lambda_{03} = \xi^{\nu}_{(0)}\xi_{\nu (3)},\nonumber\\  
\lambda_{33} &=& \xi^{\nu}_{(3)}\xi_{\nu (3)},\;\;\;
\tau^{2} = - 2 (\lambda_{00} \lambda_{33} - \lambda_{03}^{2}).
\end{eqnarray}
Applying the above definition to the present case, we find
\begin{equation}
\mu = \frac{\sigma}{4\sigma^{2} - 2\sigma + 1},
\end{equation}
which shows that when $0 < \sigma \ll 1$, we have $\mu \approx \sigma$.
This is consistent with its Newtonian limit \cite{GH1969}. However as
$\sigma$ increases, $\mu$ monotonically increases until $\sigma =
\frac{1}{2}$ where it reaches its maximum $\mu = \mu_{max.} =
\frac{1}{2}$ in the chosen units.  This maximal value is
twice that obtained in \cite{LO1980} for a particular
dust source. Thus the results obtained here and those in \cite{LO1980}
strongly suggest that the linear mass of static cylinders has an upper
limit, although with particular solutions it is difficult to find the exact
limit.  After the point $\sigma =
\frac{1}{2}$, $\mu$ starts to decrease as $\sigma$ continuously
increases. Therefore, $\sigma = \frac{1}{2}$ is a turning point. This
point is also the critical point that separates ``normal" gauge cosmic
strings from the supermassive ones. In \cite{LG1989} it is shown that
the spacetime of a gauge cosmic string asymptotically approaches the
one of Eq.(3) with $\sigma = 0$ and $C > 1$, when the mass per unit
length of the string is very low, while in the opposite limit, the
spacetime asymptotically approaches the one of Eq.(3) with $\sigma =
1$. In between these two states there exists a critical state that
separates the normal strings from the supermassive ones, and this
critical state is exactly given by the LC solution (3) with $\sigma =
\frac{1}{2}$!

On the other hand, using the definition of Israel \cite{Israel1977},
we find that 
\begin{equation}
\mu_{Israel} = \int{(\rho + p_{z} + p_{\varphi})\delta(r - r_{0})
\sqrt{g} dr d\varphi} = \sigma, 
\end{equation}
while the ``inertial" mass per unit length of Marder 
\cite{Marder1958} yields
\begin{equation}
\mu_{Marder} = \int{\rho \delta(r - r_{0})
\sqrt{g_{(2)}} dr d\varphi} = \frac{\sigma}{2(4\sigma^{2} - 
2\sigma + 1)}, 
\end{equation}
where $g_{(2)}$ is the determinant of the induced metric on the
2-surface $S$ defined by $t, z = Const.$ Clearly, in the present case
Marder's definition does not give the correct Newtonian limit, but
Israel's does. Moreover, Eq.(17) has a maximum at $\sigma = 1/2$,
while Eq.(16) does not.

\section{Conclusions}

In this paper, we have shown that the LC solution with $0 \le \sigma
\le 1$ can be produced by physically realistic cylindrical sources, and
that the mass per unit length of the cylinder, $\mu$, depends only on
the parameter $\sigma$, and is given explicitly by Eq.(15). When $0 <
\sigma \ll 1$, we have $\mu \approx \sigma$, which is consistent with
its Newtonian limit. As $\sigma$ increases, $\mu$ is monotonically
increasing until $\sigma = \frac{1}{2}$, where it reaches its maximum
$\mu = \mu_{max.} = \frac{1}{2}$. This explains why no horizons exist
in the LC solution. When $ \sigma < 0$, Eq.(12) shows that the shell
does not satisfy any of the energy conditions, and the mass of the
shell is negative. Nonetheless this is also consistent with its
Newtonian limit \cite{GH1969,BM1991}. But when $\sigma \ll 0$ one has
to consider the above results with great caution since once the energy
conditions are violated  we could, in principle, have any kind of
sources.  In particular, when $\sigma = - \frac{1}{2}$, Eq.(7) shows
that $\rho, p_{z}$ and $p_{\varphi}$ are all negative. Since the
solution with $\sigma = - \frac{1}{2}$ is locally isometric to Taub's
plane solution \cite{Taub1951}, can we conclude that the source for the
Taub solution also has negative mass? This is a question that is still
under our investigation.

\section*{Acknowledgments}

One of the authors (AZW) would like to thank J. Skea for reading
carefully the manuscript and computer help.  The financial assistance
from CNPq is gratefully acknowledged.


\end{document}